%% file: example.tex
\documentclass[submission,copyright,creativecommons]{eptcs}
\providecommand{\event}{ICLP DC 2022} 
\usepackage[table]{xcolor}
\usepackage{breakurl}             
\usepackage{underscore}           
\usepackage[inline]{enumitem}
\usepackage{url}
\usepackage{amsmath}
\usepackage{graphicx}
\usepackage{multirow}
\usepackage{tikz}
\usepackage[title]{appendix}
\usepackage{listings}
\usepackage{footmisc}
\usepackage{bookmark}
\usepackage{csvsimple}
\usepackage{libertine}
\usepackage{times}
\usepackage{soul}
\usepackage{mathtools}
\usepackage{url}
\usepackage[utf8]{inputenc}
\usepackage[style=base]{caption}
\usepackage{xspace}
\usepackage{amsthm}
\usepackage{fancyvrb}
\usepackage[ruled,vlined,linesnumbered]{algorithm2e}
\usepackage{algorithmic}
\usepackage{array}
\usepackage{subcaption}
\usepackage{booktabs,siunitx}

\newcommand{\dnot}{\ensuremath{\mathit{not}\,}}
\title{A Model-Oriented Approach for Lifting Symmetries in Answer Set Programming\thanks{This PhD work is conducted under the supervision of professors Martin Gebser and Konstantin Schekotihin.}}
\author{Alice Tarzariol  
\institute{Alpen-Adria-Universität\\ Klagenfurt, Austria}
\email{alice.tarzariol@aau.at }}
\def\titlerunning{A Model-Oriented Approach for Lifting Symmetries in ASP}
\def\authorrunning{A. Tarzariol}
\begin{document}
\maketitle

\begin{abstract}
  When solving combinatorial problems, pruning symmetric solution candidates from the search space is essential. Most of the existing approaches are instance-specific and focus on the automatic computation of Symmetry Breaking Constraints (SBCs) for each given problem instance. However, the application of such approaches to large-scale instances or advanced problem encodings might be problematic since the computed SBCs are propositional and, therefore, can neither be meaningfully interpreted nor transferred to other instances. As a result, a time-consuming recomputation of SBCs must be done before every invocation of a solver. 
  To overcome these limitations, we introduce a new model-oriented approach for Answer Set Programming that lifts the SBCs of small problem instances into a set of interpretable first-order constraints using a form of machine learning called Inductive Logic Programming. After targeting simple combinatorial problems, we aim to extend our method to be applied also for advanced decision and optimization problems.
  
\end{abstract}
\section{Introduction and Problem Description} \label{ch-1}
A common approach for solving combinatorial problems is modelling them using declarative programming paradigms, e.g., Answer Set Programming (ASP) \cite{gekakasc12a,gellif88b,breitr11a}.
In general, defining such models is relatively simple, and the obtained encodings are easy to understand. However, although correct, a trivial encoding might become useless because of its performance when solving non-trivial instances. 
Indeed, the solving phase turns infeasible when the size of input instances and, correspondingly, the number of possible solution candidates start to grow~\cite{dogalemurisc16a}.
In many cases, these candidates are symmetric, i.e., one candidate can easily be obtained from another by renaming constants.
Therefore, the ability to encode \textit{Symmetry Breaking Constraints} (SBCs) in a program becomes an essential skill for programmers as they prune a consistent part of the search space. 
Nevertheless, identifying symmetric solutions and formulating constraints that remove only them might be a time-consuming and challenging task.  
As a result, various tools emerged for avoiding the computation of symmetric solutions. 
A popular approach consists in automatically detecting and introducing a set of SBCs using properties of permutation groups \cite{sakallah09a}. The system \textsc{sbass} \cite{drtiwa10a} implements this type of approach for ground ASP programs. 

Unfortunately, the computational advantages derived from \textsc{sbass} or, more generally, from any \emph{instance-specific} symmetry breaking approach, do not carry forward to large-scale instances or advanced encodings. Indeed, instance-specific approaches often require as much time as it takes to solve the original problem.
Moreover, ground SBCs generated approaches are 
\begin{enumerate*}[label=\emph{(\roman*)}]
  \item not transferable, since the knowledge obtained is limited to a single instance;
    \item usually hard to interpret and comprehend because they are not expressed with a symbolic representation;
    \item derived from permutation group generators, whose computation is itself a combinatorial problem; and
    \item often redundant and might result in a degradation of the solving performance.
\end{enumerate*}
In particular, when solving instances sharing similar characteristics, the identified symmetries follow the same structure for the whole set. Thus, the instance-specific approaches require extra computation time to identify the same symmetries applied to every single instance.
Assuming we consider a combinatorial ASP program and a distribution of similar problem instances, in our work, we aim to overcome the limitations of instance-specific approaches by lifting ground symmetries using a form of machine learning called Inductive Logic Programming (ILP) \cite{crdumu20a}. The resulting first-order constraints can be applied to any instance drawn from the considered distribution, and they should speed up the identification of satisfiable/unsatisfiable instances.

\section{Background and Existing Literature}
In this section, we will briefly introduce the concepts considered in our work, i.e., Answer Set Programming, Inductive Logic Programming and Symmetry Breaking techniques.
\subsection{Answer Set Programming}
ASP is a declarative programming paradigm that applies non-monotonic reasoning and relies on the stable model semantics \cite{gellif91a}. 
Over the past decades, it has attracted considerable interest thanks to its elegant syntax, expressiveness, and efficient system implementations. It showed promising results in numerous domains, including industrial, robotics, or biomedical applications \cite{ergele16a}. 

\paragraph{Syntax.} 
An ASP program $P$ is a set of \textit{rules}~$r$ of the form: 
\begin{equation*}
  a_0 \gets a_1, \dots, a_m, \dnot a_{m+1}, \dots, \dnot a_n
\end{equation*}
where $\mathit{not}$ stands for \textit{default negation} and $a_i$, for $0\leq i \leq n$, are atoms. 
An \textit{atom} is an expression of the form $p(\overline{t})$, where $p$ is a predicate, $\overline{t}$ is a possibly empty vector of terms, and the predicate $\bot$ (with an empty vector of terms) represents the constant \textit{false}.
Each \textit{term} $t$ in $\overline{t}$ is either a variable or a constant.
A \textit{literal} $l$ is 
an atom $a_i$ (positive) or its negation $\dnot a_i$ (negative). 
The atom $a_0$ is the \textit{head} of a rule~$r$,
denoted by $H(r)=a_0$,
and the \textit{body} of~$r$ includes the positive or negative, respectively,
body atoms $B^+(r) = \{a_1, \dots, a_m\}$ and $B^-(r) = \{a_{m+1}, \dots, a_n\}$.
%
A rule~$r$ is called a \textit{fact} if $B^+(r)\cup B^-(r)=\emptyset$, and a \textit{constraint} if $H(r) = \bot$.

\paragraph{Semantics.}
The semantics of an ASP program $P$ is given in terms of its \textit{ground instantiation} $P_{\mathit{grd}}$, which is obtained by
replacing each rule $r\in\nolinebreak P$ with its instances obtained by substituting the variables in $r$ by constants occurring in $P$.
Then, an \textit{interpretation} $\mathcal{I}$ is a set of (\textit{true}) ground atoms occurring in $P_{\mathit{grd}}$ that does not contain~$\bot$. 
An interpretation $\mathcal{I}$ \textit{satisfies} a rule $r \in P_{\mathit{grd}}$ if $B^+(r) \subseteq \mathcal{I}$ and $B^-(r) \cap \mathcal{I}=\emptyset$ imply $H(r) \in \mathcal{I}$, and
$\mathcal{I}$ is a \textit{model} of $P$ if it satisfies all rules $r\in P_{\mathit{grd}}$. 
A model $\mathcal{I}$ of~$P$ is \textit{stable} if it is a subset-minimal model of the reduct $\{H(r) \gets B^+(r) \mid r \in P_{\mathit{grd}}, B^-(r) \cap \mathcal{I} = \emptyset\}$, and we denote the set of all stable models, also called answer sets, of $P$ by $\mathit{AS}(P)$.

\subsection{Inductive Logic Programming}
ILP is a form of machine learning whose goal is to learn a logic program that explains a set of observations in the context of some pre-existing knowledge.
The most expressive ILP system for ASP is \emph{Inductive Learning of Answer Set Programs} (\textsc{ilasp}) \cite{larubr20b,ilasp}, which can be used to solve a variety of ILP tasks.
A learning task $\langle B,E^+,E^-,H_M \rangle$ is defined by four elements: a background knowledge $B$, a set of positive and negative examples, respectively $E^+$ and $E^-$, and lastly a hypothesis space $H_M$, which defines the rules that can be learned. 
Each example $e \in E^+ \cup E^-$ is a pair $\langle e_{\mathit{pi}}, C\rangle$ called \textit{Context Dependent Partial Interpretation}, where 
\begin{enumerate*}[label=\emph{(\roman*)}]
  \item $e_{\mathit{pi}}$ is a \textit{Partial Interpretation} defined as pair of sets of atoms $\langle T, F\rangle$, called \textit{inclusions} ($T$) and  \textit{exclusions} ($F$), respectively, and
  \item $C$ is an ASP program defining the \textit{context} of $e_{\mathit{pi}}$.
\end{enumerate*}
Given a (total) interpretation $\mathcal{I}$ of a program $P$ and a partial interpretation $e_{\mathit{pi}}$, we say that $\mathcal{I}$ \textit{extends} $e_{\mathit{pi}}$ if $ T \subseteq \mathcal{I}$ and $F \cap \mathcal{I} = \emptyset$.
Given an ASP program $P$, an interpretation $\mathcal{I}$, and an example $e=\langle e_{\mathit{pi}}, C\rangle$, we say that  $\mathcal{I}$ is an \textit{accepting answer set} of $e$ with respect to $P$ if $\mathcal{I} \in \mathit{AS}(P \cup C)$ such that $\mathcal{I}$ extends $e_{\mathit{pi}}$.

Each hypothesis $H \subseteq H_M$ learned by \textsc{ilasp} must respect the following criteria: 
\begin{enumerate*}[label=\emph{(\roman*)}]
  \item for each positive example $e \in E^+$, there is some accepting answer set of $e$ with respect to $B\cup H$; and 
  \item for any negative example $e \in E^-$, there is no accepting answer set of $e$ with respect to $B\cup H$.
\end{enumerate*} 
If multiple hypotheses satisfy the conditions, the system returns one of those with the lowest cost. By default, the cost $c_{r}$ of each rule $r \in H_M$ corresponds to its number of literals \cite{larubr14a}; however, the user can define a custom scoring function for defining the rule costs.
\textsc{Ilasp} allows to define learning tasks with noisy examples \cite{larubr18b}.
With this setting, if an example $e$ is not covered, i.e., there is an accepting answer set for $e$ if it is negative, or none if $e$ is positive, the corresponding weight 
is counted as a penalty. If no dedicated weight is specified, the example's weight is infinite, thus forcing the system to cover the example. Therefore, the learning task becomes an optimization problem with two goals: minimize the cost of $H$ and minimize the total penalties for the uncovered examples. 
In our work, we use the most recent version of \textsc{ilasp} (v4.1.2), which implements the search approach \textit{Conflict Driven ILP} \cite{law21a}.

\subsection{Symmetry Breaking}
Modern symmetry breaking approaches can be split into two families: \emph{instance-specific} and \emph{model-oriented} approaches \cite{sakallah09a,walsh12a}.
The former identify symmetries for a particular instance at hand by obtaining a ground program, computing ground SBCs, composing a new extended program, and solving it~\cite{DBLP:conf/cp/Puget05,DBLP:journals/constraints/CohenJJPS06,drtiwa10a}. The system \textsc{sbass} \cite{drtiwa10a} implements this type of approach for ground ASP programs. 
As mentioned in Section \ref{ch-1}, when applied to large-scale instances or advanced encodings, the \emph{instance-specific} symmetry breaking approaches may struggle to compute the symmetries in a reasonable time, or the resulting SBCs are redundant, leading to more drawbacks than benefits for the solver.
Moreover, the ground SBCs are often difficult to understand as they are not expressed in symbolic representation; e.g., the SBCs produced by \textsc{sbass} are represented in \textsc{smodels} format \cite{lparseManual}.

In contrast, \emph{model-oriented} approaches aim to find general SBCs that depend less on a particular instance. The method presented in \cite{debobrde16a} uses local domain symmetries of a given first-order theory. SBCs are generated by identifying argument positions in atoms of a formula that comprise object variables defined over the same subset of a domain given in the input. As a result, the computation of lexicographical SBCs is very fast. However, the method considers each first-order formula separately and cannot reliably remove symmetric solutions, as it requires the analysis of several formulas at once.
The method of \cite{DBLP:conf/cpaior/MearsBWD08} computes SBCs by generating small instances of parametrized constraint programs, and then finds candidate symmetries using \textsc{saucy} \cite{cokasama13a,saucy} -- a graph automorphism detection tool. Next, the algorithm removes all candidate symmetries that are valid only for some of the generated examples as well as those that cannot be proven to be parametrized symmetries using heuristic graph-based techniques. This approach can be seen as a simplified learning procedure that utilizes only negative examples represented by the generated SBCs.

\section{Current Research}
This section describes the research goals targeted for my PhD, and the current state of my work.
\subsection{Goal of the Research}
To the best of our knowledge, currently there are no \emph{model-oriented} systems that lift ground SBCs for ASP programs. 
Therefore, we aim to introduce a novel \emph{model-oriented} method that generalizes the process of discarding redundant solution candidates for ASP instances of a target domain using ILP. 
More precisely, we identify and lift SBCs of small problem-instances, obtaining a set of interpretable first-order constraints. Such constraints cut the search space while preserving the satisfiability of a problem for the considered instance distribution, which improves the solving performance, especially in the case of unsatisfiability. After targeting simple combinatorial problems, we aim to extend our method to be applied also for advanced decision and optimization problems.

The research goals of this work are the following:
\begin{itemize}[]
  \item[\textbf{RG 1}] Given an ASP combinatorial program and a target instances distribution, define a learning framework capable of obtaining first-order constraints that speed up the solving of satisfiable and unsatisfiable instances.
  
  \item[\textbf{RG 2}] Develop an approach capable of applying the learning framework 
  iteratively.
  

  \item[\textbf{RG 3}] Investigate how the framework 
  can be extended to enable learning first-order constraints for advanced combinatorial problems.
  
  \item[\textbf{RG 4}] Design and implement systems that automate parameter selection for the framework 
  for guiding the learning of first-order constraints that speed up solving.

  \item[\textbf{RG 5}] Extend the expressiveness of the learning framework 
  to analyse the symmetries on optimization problems.
  \end{itemize}
  For the research goal \textbf{RG 1}, we assume that the instances analyzed for a given ASP combinatorial problem $P$ follow a specific distribution. 
  Moreover, we can easily provide a set of simple instances (i.e., such that the total number of solutions can be managed by \textsc{sbass}, \textsc{clingo}\footnote{In our work, to find the solutions of ASP programs, we use the system \textsc{clingo}, consisting of the grounding and solving components \textsc{gringo} and \textsc{clasp}.} and \textsc{ilasp}) that entail the symmetries of the whole target distribution. 
  Our framework defines the set of examples of an ILP task such that \textsc{ilasp} learns 
  first-order constraints that remove symmetric solutions while preserving the satisfiability of the instances in the considered distribution. To do so, the framework relies on \textsc{sbass} to compute the SBCs of a set of small, satisfiable, and representative problem-instances, identified with $S$. 
More precisely, for each instance $i\in S$, we find its symmetries expressed as a set of irredundant generators, $\mathit{IG(i)}$, and we enumerate the set of its answer sets, $AS(i)$. Then, for each interpretation~$\mathcal{I} \in AS(i)$, we define an example where $\mathit{IG(i)} \cap \mathcal{I}$ and $\mathit{IG(i)} \setminus \mathcal{I}$ are the inclusions and exclusions of a partial interpretation and using $i$ as context. If $\mathcal{I}$ is dominated, i.e., $\mathcal{I}$ can be mapped to a lexicographically smaller, symmetric answer set by means of some irredundant generator in $\mathit{IG(i)}$, the example is labelled as negative, otherwise, positive.
Together with $S$, we use another set of instances $\mathit{Gen}$, where each $g \in \mathit{Gen}$ defines a single positive example with empty inclusions and exclusions and $g$ as context. These examples guarantee that the learned constraints generalize for the target distribution since they force the constraints to preserve some solution for each $g \in \mathit{Gen}$. 
The two sets of instances, $S$ and $Gen$, and the language bias for the ILP task are defined by the user,  while $P$ is used as background knowledge.  
After the learning phase, a validation set of satisfiable instances, $V$, is used to check whether the learn constraints in $ABK$ are correct. If there is a satisfiable instance $v \in V$ for which no solution is found, we discard the learned constraints and add $v$ in $Gen$. Subsequently, we rerun our framework and repeat the procedure until all the instances in the validation set are satisfiable.

The research goal \textbf{RG 2} consists of identifying techniques that can speed up the learning phase for the tasks analysed in \textbf{RG 1} as the learning time represent a critical aspect on ILP \cite{cropdum20a}. To do so, we define a procedure that applies our framework iteratively to learn the first-order constraints incrementally.  
More precisely, we outline a criterion for splitting the framework inputs to create sub-learning tasks. This approach speeds up the computation of first-order constraints, especially when the program contains symmetries independent from each others. To do so, we introduce an auxiliary ASP file called \textit{Active Background Knowledge} or $ABK$, containing the constraints learned so far. By including $ABK$ in the background knowledge, we can rerun our framework taking into account the constraints previously learned.

The current definition of our framework yields a number of examples proportional to the number of solutions for each problem-instance in $S$. Therefore, if it gets difficult to compute all the solutions for an instance in~$S$ to analyze, the resulting formulation of the ILP task to learn constraints can become prohibitive.
The research goal \textbf{RG 3} consists of overcoming the limitations of the current framework, in order to apply it to advanced combinatorial problems. 
With the term ``advanced'', we refer to problems whose solutions rely on atoms of multi-dimensional instead of just unary predicates,
so that there might be no trivial instances to analyse.
An example of this kind of problems is the \textit{Partner Units Problem} (PUP) \cite{DBLP:conf/cpaior/AschingerDFGJRT11,DBLP:journals/jcss/TeppanFG16}, which is an abstract representation of configuration problems occurring in railway safety or building security systems.
Considering the smallest PUP instance representing a class of building security systems named \textit{double} by \cite{DBLP:conf/cpaior/AschingerDFGJRT11}, \textsc{clingo} finds $145368$ solutions, $98.9\%$ of which can be identified as symmetric by \textsc{sbass} (for instance, by renaming the units of a solution). Thus, the enumeration of symmetries for PUP instances is problematic, even for the smallest and simplest ones.
To overcome this problem, we need to revise the framework's approach such that it manages to cope with any number of solutions of the analyzed instances, for example, by sampling a subset of answer sets. To be effective, the sample size must be small while containing an adequate number of positive and negative examples. Besides, two further limitations need to be addressed concerning \textsc{ilasp}'s searching technique. The former is the inefficiency of the default \textsc{ilasp}'s conflict analysis techniques\footnote{A key component of \textit{Conflict-Driven ILP}.} when applied to positive examples producing many solutions (i.e., the examples generated from the PUP instances in $Gen$). The second issues concerns the optimal criterion for the learned hypothesis, which considers only the length of the constraints and not the nature of the predicates. To overcome both issues, we aim to devise a specific conflict analysis technique that exploit the nature of the learning task (namely, constraints learning) and a custom scoring function for \textsc{ilasp} that provides further information for learning efficient\footnote{Namely, constraints with a limited number of variables and, possibly, containing some predicates that are simplified during grounding.} constraints. 


For the research goal \textbf{RG 4}, we aim to identify appropriate inputs to our framework automatically.
So far the inputs must be chosen by the user, however, we would like to provide guidelines or automate the process of selecting the framework inputs. 
First, the selection method of our framework needs to assess the properties of candidate inputs. Then, the method should determine parameters leading to correct and performant first-order constraints. That is, it aims to find constraints preserving at least one solution for satisfiable instances and cutting down the solving time for unsatisfiable instances. 

Lastly, the research goal \textbf{RG 5} is to extend the applicability of our framework to optimization problems.
The tool \textsc{sbass}, used in our framework, can analyse most of the ASP rules as normal rules, choice rules, aggregates, and hard constraints; however, it does not support weak constraints. Therefore, we aim to define a reduction from programs with weak constraints to an equivalent representation that can be processed by \textsc{sbass}. 
More specifically, we aim to introduce new (normal) rules that respect the symmetries of the optimisation rules.
As a consequence, when running \textsc{sbass} on the extended program, we get a finer partition of the solutions. Namely, it could be that two solutions, which were considered symmetric from the analysis of the original program, can be identified as non-symmetric after the introduction of the new rules.

\subsection{Results Accomplished}
We devised and implemented the learning framework of \textbf{RG 1} and a rough idea of \textbf{RG 2} in a conference paper \cite{tagesc21a}; subsequently, we formalised the method to split the learning task in a journal paper \cite{mljournal}. 
We applied the framework to simple ASP programs, namely, the pigeon-hole problem and two its extensions that consider also the assignments of colors and owners. Moreover, we tested the house-configuration problem \cite{DBLP:conf/confws/FriedrichRFHSS11}. For all the addressed problems, we suggested some guidelines to define the framework inputs, $S$, $Gen$, $H_M$, and $ABK$. 
Table \ref{table:hc} contains the solving times for the house-configuration problem; 
the satisfiable instances are shown in grey rows, while the white rows contain unsatisfiable instances. The column \textbf{\textsc{base}} refers to \textsc{clingo} (v5.5.0)
run on the original encoding, while \textbf{\textsc{abk}} reports results for the original encoding augmented with first-order constraints learned with our framework.
The time required by \textsc{sbass} to compute ground SBCs is given in the corresponding column, and \textbf{\textsc{clasp}}$^\pi$ provides the solving time obtained with these ground SBCs. Therefore, the total time required for the online usage of \textsc{sbass} is the sum of \textbf{\textsc{base}} and \textbf{\textsc{clasp}}$^\pi$.
Runs that did not finish within the time limit of 900 seconds are indicated by TO entries.

The running times in the table show the limits of \textsc{sbass} both in the pre-solving phase, when computing the symmetries (obtaining a timeout for all the satisfiable instances), and when solving a program extended with redundant constraints (the performance degradation is visible with the instance \texttt{p5-c6-t13}).  
The \textbf{\textsc{base}} encoding is quicker than \textbf{\textsc{sbass}+}\textbf{\textsc{clasp}}$^\pi$ to solve satisfiable instances, although it takes considerably longer for unsatisfiable ones.
On the other hand, the first-order constraints learned with our framework helped the search for satisfiable and, especially, unsatisfiable instances.
Similar results have also been observed for the pigeon-hole problems analysed. The repository containing the implementation and complete experiments can be found at the following link: \url{https://github.com/prosysscience/Symmetry_Breaking_with_ILP/tree/extended}
\begin{table}[t]
  \centering
  \rowcolors{2}{gray!25}{white}
  \setlength{\tabcolsep}{7.5pt}
  \resizebox{8cm}{!}{
  \csvloop{
  file=HouseConfiguration.csv,
  head to column names,
  before reading=\centering\sisetup{table-number-alignment=center},
  tabular={lrrrr},
  table head=\toprule  & \textbf{\textsc{ABK}} & \textbf{BASE} &  \textbf{SBASS} & $\mathbf{CLASP^\pi}$\\\midrule,
  command=\Instance & \ABKfull & \BASE  &  \SBASS & \Clasp,
  table foot=\bottomrule}}
  \caption{Runtime in seconds for house-configuration problem.}
  \label{table:hc}
\end{table}%

We addressed the extension mentioned in \textbf{RG 3} in a paper presented at ICLP 2022 \cite{iclp}. In this paper, we revised several parts of our framework in order to target PUP instances supplied by \cite{DBLP:conf/cpaior/AschingerDFGJRT11}, studying the \textit{double}, 
\textit{doublev}, 
and \textit{triple} 
instance collections. Instances of the same type represent buildings of similar topology with scaling parameters that follow a common distribution. Although the benchmark instances are synthetic, they represent a relevant configuration problem concerning safety and security issues in public buildings, like administration offices or museums. In addition, the scalable synthetic benchmarks are easy to generate and analyze. 

Table \ref{table:double} contains the solving time for the PUP instances in \textit{double}, and the table follows the same structure as the previous, but with a timeout of 600 seconds.
Moreover, it also considers the computational time obtained  by running \textsc{clingo} on the advanced encoding \textbf{\textsc{symm}}\footnote{The encoding used for \textbf{\textsc{symm}} is taken from the paper \cite{dogalemurisc16a}, where it is called \textsc{ENC2}. From the same paper, we also take the basic encoding \textsc{ENC1} and use it as \textbf{\textsc{base}}.} which incorporates hand-crafted static symmetry breaking as well as an ordered representation \cite{crabak94a} of assigned units.
From the table, we can observe that \textbf{\textsc{symm}} leads
to more robust \textsc{clingo} performance than the simpler \textbf{\textsc{base}}
encoding, and the ground SBCs computed from \textsc{sbass} (obtained by summing the time in \textbf{\textsc{sbass}} with \textbf{\textsc{clasp}}$^\pi$).
Moreover, when comparing \textbf{\textsc{symm}} to \textbf{\textsc{abk}},
we observe further significant performance improvements thanks to our approach, particularly on the unsatisfiable instances.
That is, the learned $\mathit{ABK}$ enables \textsc{clingo}
to solve the considered PUP instances and
efficiently prunes the search space, which must be fully explored in case of unsatisfiability.
Similar results have been observed also for the other two type of instances, \textit{doublev} and \textit{triple}. The repository containing the implementation and complete experiments can be found at the following link: \url{https://github.com/prosysscience/Symmetry_Breaking_with_ILP/tree/pup}

\begin{table}[tb]
      \centering
      \rowcolors{2}{gray!25}{white}
      \setlength{\tabcolsep}{7.5pt}
      \resizebox{9cm}{!}{
      \csvloop{
      file= PUPdouble.csv,
      head to column names,
      before reading=\centering\sisetup{table-number-alignment=center},
      tabular={lrrrrr},
      table head=\toprule  & \textbf{ABK} & \textbf{SYMM} & \textbf{BASE} & \textbf{SBASS} & $\mathbf{CLASP^\pi}$\\\midrule,
      command=\Instance  & \ABK & \ENC & \BASE & \SBASS & \CLASP,
      table foot=\bottomrule}}
      \caption{Runtimes for PUP double}
      \label{table:double}
\end{table}

\subsection{Open Issues and Expected Achievements}
The research goals that we still need to tackle are \textbf{RG 4} and \textbf{RG 5}. For the former, we would like to help the user on deciding the elements in $S$ and $Gen$, and automatically identify a set of constraints which performs relatively fast, while preserving the satisfiability of the target instances. A more ambitious target that can be developed for this research goal is the identification of predicates to use in the language bias to learn the constraints. 
Moreover, from the experiments with the PUP instances, we observed that the labelling instance impacted the symmetries identified by \textsc{sbass}. Thus, we hope 
to introduce automatic (re-)labeling schemes for constants appearing in instances to exploit common problem structure in a less input-specific way.
For \textbf{RG 5} we aim to extend the applicability of our framework to programs containing weak constraints. Targeting optimization problems can lead to relevant results as optimization involves solving unsatisfiable subproblem(s) on attempting (and failing) to improve an optimal answer set, where symmetry breaking is particularly crucial for the performance.

\bibliographystyle{eptcs}
\bibliography{mybib,krr,procs}

\input{appendix}
\end{document}

%% file: appendix.tex
\newpage
\begin{appendices}
\section{Inductive Learning from Symmetries - Example}
Here we will illustrate an example of our ILP framework; for more details, see \cite{mljournal}. 
Let us consider the pigeon-hole problem,
which is about checking whether $p$ pigeons can be placed into $h$ holes
such that each hole contains at most one pigeon.
An encoding in ASP of this problem is:
\begin{Verbatim}[fontsize=\small]
 pigeon(X-1) :- pigeon(X), X > 1.
 hole(X-1) :- hole(X), X > 1.
 {p2h(P,H) : hole(H)} = 1 :- pigeon(P).
 :- p2h(P1,H), p2h(P2,H), P1 != P2.
\end{Verbatim}
It takes as input the ground facts \texttt{pigeon(}$p$\texttt{).} and \texttt{hole(}$h$\texttt{).} For example, solving the instance with $p=3$ and $h=3$ leads to six answer sets: 
\begin{lstlisting}
  $\mathit{AS}_1$ = {p2h(1,1), p2h(2,2), p2h(3,3)} = 100010001
  $\mathit{AS}_2$ = {p2h(1,1), p2h(2,3), p2h(3,2)} = 010100001
  $\mathit{AS}_3$ = {p2h(1,2), p2h(2,1), p2h(3,3)} = 100001010
  $\mathit{AS}_4$ = {p2h(1,2), p2h(2,3), p2h(3,1)} = 001100010
  $\mathit{AS}_5$ = {p2h(1,3), p2h(2,1), p2h(3,2)} = 010001100
  $\mathit{AS}_6$ = {p2h(1,3), p2h(2,2), p2h(3,1)} = 001010100
\end{lstlisting}
where the binary integer given on the right corresponds to the value that will be considered for the lexicographic order. 
Using \textsc{sbass} with this instance produces the following set of generators:
\begin{lstlisting}
  $\pi_1 \; = \; \big(\,$p2h(3,2) p2h(3,3)$\,\big)$ $\big(\,$p2h(2,2) p2h(2,3)$\,\big)$ $\big(\,$p2h(1,2) p2h(1,3)$\,\big)$
  $\pi_2 \; = \;  \big(\,$p2h(3,1) p2h(3,3)$\,\big)$ $\big(\,$p2h(2,1) p2h(2,3)$\,\big)$ $\big(\,$p2h(1,1) p2h(1,3)$\,\big)$ 
  $\pi_3 \; = \; \big(\,$p2h(2,3) p2h(3,3)$\,\big)$ $\big(\,$p2h(2,2) p2h(3,2)$\,\big)$ $\big(\,$p2h(2,1) p2h(3,1)$\,\big)$ 
  $\pi_4 \; = \;  \big(\,$p2h(1,1) p2h(3,3)$\,\big)$ $\big(\,$p2h(2,1) p2h(2,3)$\,\big)$ $\big(\,$p2h(1,3) p2h(3,1)$\,\big)$
     $\quad\big(\,$p2h(1,2) p2h(3,2)$\,\big)$
\end{lstlisting}
Applying a generator to an answer set returns a symmetric solution. For example, $\pi_1(\mathit{AS}_6) = \mathit{AS}_4$.
For each answer set $AS_i$, we apply all the generators to it and check whether there is a generator $\pi_j$ such that $AS_i \geq \pi_j(\mathit{AS}_i)$. If there exists such $\pi_j$, then $AS_i$ will define a negative example, otherwise a positive one. 

As a result, we create one positive example with $\mathit{AS}_6$ (since it is the only answer set that is not mapped into a smaller interpretation) and five negative examples with the other answer sets. The resulting ILP task is as follows:
  \begin{Verbatim}[fontsize=\small]
 %% Input encoding 
 pigeon(X-1) :- pigeon(X), X > 1.
 hole(X-1) :- hole(X), X > 1.
 {p2h(P,H) : hole(H)} = 1 :- pigeon(P).
 :- p2h(P1,H), p2h(P2,H), P1 != P2.
  
 %% Active Background Knowledge
 lessThan(X,Y) :- pigeon(X), pigeon(Y), X < Y.
 lessThan(X,Y) :- hole(X), hole(Y), X < Y.
 maxpigeon(X) :- pigeon(X), not pigeon(X+1).
 maxhole(X) :- hole(X), not hole(X+1).

 %% Negative examples
 #neg(id1@100, {p2h(2,3), p2h(1,2), p2h(3,1)}, 
  {p2h(2,1), p2h(1,1), p2h(3,3), p2h(1,3), p2h(3,2), p2h(2,2)},
  {pigeon(3). hole(3).}).
 #neg(id3@100, {p2h(2,1), p2h(3,2), p2h(1,3)},
  {p2h(1,1), p2h(3,3), p2h(3,1), p2h(2,2), p2h(2,3), p2h(1,2)},
  {pigeon(3). hole(3).}).
 #neg(id4@100, {p2h(2,3), p2h(1,1), p2h(3,2)},
  {p2h(2,1), p2h(3,3), p2h(3,1), p2h(1,3), p2h(2,2), p2h(1,2)},
  {pigeon(3). hole(3).}).
 #neg(id5@100, {p2h(2,1), p2h(3,3), p2h(1,2)},
  {p2h(1,1), p2h(3,1), p2h(1,3), p2h(3,2), p2h(2,3), p2h(2,2)}, 
  {pigeon(3). hole(3).}).
 #neg(id6@100, {p2h(1,1), p2h(3,3), p2h(2,2)},
  {p2h(2,1), p2h(3,1), p2h(1,3), p2h(3,2), p2h(2,3), p2h(1,2)},
  {pigeon(3). hole(3).}).
   
 %% Positive example 
 #pos(id2, {p2h(3,1), p2h(2,2), p2h(1,3)}, {}, 
  {pigeon(3). hole(3).}).
 
 %% Language bias
 #modeb(2,p2h(var(pigeon),var(hole))).
 #modeb(2,pigeon(var(pigeon))).
 #modeb(2,hole(var(hole))).
 #modeb(1,maxhole(var(hole))).
 #modeb(1,maxpigeon(var(pigeon))).
 #modeb(2,lessThan(var(hole),var(hole)),(anti_reflexive)).
 #modeb(2,lessThan(var(pigeon),var(pigeon)),(anti_reflexive)).
 #modeb(2,lessThan(var(hole),var(pigeon))).
 #modeb(2,lessThan(var(pigeon),var(hole))).
  \end{Verbatim}
 
  After running \textsc{ilasp}, the learned first-order constraints are:
  \begin{Verbatim}[fontsize=\small]
 :- p2h(X,Y), lessThan(Z,Y), maxpigeon(X).
  % do not assign the pigeon with the max label to a hole
  % other than the first one
 :- p2h(X,Y), lessThan(X,Y), lessThan(Y,Z).
  % for all but the last hole, do not assign a pigeon with 
  % a smaller label to the hole
   \end{Verbatim}
\end{appendices}